\documentclass[final,11pt,1p,times]{elsarticle}

\journal{Journal of Parallel and Distributed Computing}
\nocite{*}
\usepackage{url}

\usepackage{blindtext}
\usepackage{graphicx}
\usepackage{subfig}
\usepackage{blindtext}
\usepackage{graphicx}
\usepackage{amsmath}
\usepackage{color}
\usepackage{algorithmicx}
\usepackage{algorithm}
\usepackage{algpseudocode}
\usepackage{caption}
\usepackage[table,xcdraw]{xcolor}
\usepackage{array,arydshln}
 \usepackage{graphicx,booktabs}
 \usepackage{longtable}
\usepackage{float}
\usepackage[normalem]{ulem}
\usepackage{rotating,multirow}
\usepackage{blindtext}

\begin{document}
	\begin{frontmatter}
		
\title{FOCAN: A Fog-supported Smart City Network Architecture for Management of Applications in the Internet of Everything Environments}

\author[label1]{Paola~G.~Vinueza~Naranjo}
\address[label1]{DIET, ``Sapienza'' University of Rome, Rome, Italy}
\ead{paola.vinueza@uniroma1.it}

\author[label2]{Zahra Pooranian}
\ead{zahra.pooranian@math.unipd.it}
\address[label2]{Department of Mathematics, University of Padua, 35131, Italy}
\cortext[cor1]{corresponding author}

\author[label1,label3]{Mohammad Shojafar\corref{cor1}}
\address[label3]{CNIT (Center of National Consortium Inter-universities in Telecommunication), Department of Electronic Engineering, ``Tor Vergata'' University of Rome, Rome, Italy}

\ead{mohammad.shojafar@uniroma1.it, mohammad.shojafar@cnit.it}

\author[label2]{Mauro Conti}
\ead{conti@math.unipd.it}

\author[label4]{Rajkumar Buyya}
\address[label4]{Cloud Computing and Distributed Systems (CLOUDS) Lab, School of Computing and Information Systems, The University of Melbourne, Melbourne, Vic. 3010, Australia}
\ead{rbuyya@unimelb.edu.au}
%
%
%
%
%

\begin{abstract}
Smart city vision brings emerging heterogeneous communication technologies such as Fog Computing (FC) together to substantially reduce the latency and energy consumption of Internet of Everything (IoE) devices running various applications. The key feature that distinguishes the FC paradigm for smart cities is that it spreads communication and computing resources over the wired/wireless access network (e.g., proximate access points and base stations) to provide resource augmentation (e.g., cyberforaging) for resource- and energy-limited wired/wireless (possibly mobile) things. Moreover, smart city applications are developed with the goal of improving the management of urban flows and allowing real-time responses to challenges that can arise in users' transactional relationships. 

Motivated by these considerations, this article presents a Fog-supported smart city network architecture called Fog Computing Architecture Network (FOCAN), a multi-tier structure in which the applications running on things jointly compute, route, and communicate with one another through the smart city environment to decrease latency and improve energy provisioning and the efficiency of services among things with different capabilities. An important concern that arises with the introduction of this framework is the need to avoid transferring data to and from distant things and instead to cover the nearest region for an IoT application. We define three types of communications between FOCAN devices --- \textit{interprimary}, \textit{primary}, and \textit{secondary} communication --- to manage applications in a way that meets the quality of service standards for the Internet of Everything. One of the main advantages of the proposed architecture is that the devices can provide the services with low energy usage and in an efficient manner. Simulation results for a selected case study demonstrate the tremendous impact of the FOCAN energy-efficient
solution on the communication performance of various types of things in smart cities. 
\end{abstract}

	\begin{keyword}		
Smart City, Internet of Everything (IoE), Fog Computing, Routing Algorithm, Computing and Communication.
\end{keyword}
\end{frontmatter}


\section{Introduction}\label{sec:1}

The \textit{smart city} concept arises from the idea of efficient use of city resources for enhancing citizens' quality of life~ \cite{17neirotti2014current}, as the pace of urban living has recently accelerated. To achieve a better quality of life, improvement of services and infrastructure in cities must be taken into account. Thanks to the revolution in information and communication technology and the power of the Internet~ \cite{23vlacheas2013enabling}, infrastructures and public services are expected to be more interactive, more accessible, and more efficient as we move towards the realization of smart cities. In this context, the emergence of the Internet of Things (IoT) paradigm strongly encourages utilization of the IoT's potential to support the smart city vision around the world. As a consequence, the smart city has emerged as one of the important IoT application drivers. Smart city IoT systems promote the concept of interrelated physical objects (\textit{things}) that are uniquely identified and distributed over broad physical areas covering an entire city. Recently, the IoT concept has taken an important step towards connecting four pillars --- things, data, process, and even people --- as the Internet of Everything (IoE). From one perspective, cities can be regarded as an aggregation of interconnected networks that make up the IoE. Hence, the IoE pillars play a significant role and work together toward the promise of our smart city vision for the future.

The IoE's generation of Big Data (BD) over a distributed environment has the potential to create data processing as well as data storage problems. One solution to address these problems is the utilization of Cloud Computing. However, some applications cannot work efficiently on the Cloud due to its inherent problems~ \cite{ifogsim}. As an example, smart city applications like health monitoring and traffic monitoring cannot tolerate the delay and latency incurred when transferring a massive amount of data to the remote Cloud Computing center and then back to the application. For this purpose, the concept of \textit{Fog Computing} (FC) recently appeared. FC extends Cloud services to the edge of the network, closer to the end user, which reduces data processing time and network traffic overhead~\cite{1varghese2017feasibility}.

The primary definition of FC was introduced by Cisco~\cite{9bonomi2012fog}. The most fundamental entity in FC, called a \textit{Fog Node} (FN), facilitates the execution of IoT applications. Basically, FC can act as an interface layer between end users / end devices and distant Cloud data centers, with the aim of satisfying mobility support, locational awareness, geodistribution, and low latency requirements for IoT applications. Since the distance between FNs and end users also varies, we propose a multi-tiered framework that does not need to transfer a vast amount of data to and from remote FNs.

We can save energy and reduce delays by avoiding transferring data or using storage resources that are too far away from the FNs by choosing FNs that are closer to the end users. Our framework also attempts to answer the following questions: What is the current state of the art in the field of smart city research, specifically, in smart city components and services? What are the main challenges that need to be addressed? This article aims to shed light on these issues and to define future research directions. Specifically, we believe that the services and components engaged in smart cities should adopt emerging technologies around the following pillars: (i)We present a generalized multi-tiered smart city architecture that utilizes FC for each device; (ii) we develop an FC-supported resource allocation model to cover device-to-device (thing-to-thing or $t2t$), device-to-FN (thing-to-FN or $t2FN$), and FN-to-FN ($FN2FN$) components; and (iii) we include various types of communications between the components and evaluate the performance of our solution on real datasets. In summary, the most important properties that distinguish this framework from other FC-supported frameworks is the utilization of resources closer to the end users based on their layer, beginning with $t2t$, then $t2FN$, and finally $FN2FN$.

The remainder of this paper is organized as follows. In Section~\ref{sec:2}, we review the most recent FC-based smart city applications and their benefits/drawbacks. In Section~\ref{sec:3}, we present a high-level view of our smart city model. Section~\ref{sec:4} presents the FC-supported smart city architecture, its hierarchical layer definitions, and their relations to the FNs. Section \ref{sec:5} presents a smart city case study for the IoE-based architecture and the numerical results attained through extensive numerical tests on a simulated scheme for an FC platform (i.e., iFogSim), with details about the test setup and the formulas utilized. Section~\ref{sec:6} describes recent issues and research directions for this problem. Finally, Section~\ref{sec:7} concludes the paper and suggests new research directions.

\section{Related Work}\label{sec:2}

Several works have been devoted to the relation between the IoT and smart cities. For instance, the Padova smart city project~\cite{22zanella2014internet} introduces characteristics of the Urban IoT system as well as the services that are required to support and implement the smart city vision. The work of~\cite{8giang2016building} shows that it is necessary to build smart city IoT applications that have distributed coordination schemes. Increasing attention is being devoted to integrating the IoT and Cloud computing in order to generate smart city applications and frameworks. The authors of~\cite{20jin2014information} provide a Cloud-based framework to create a smart city through the capabilities of the IoT. 

We have discussed some drawbacks of utilizing Cloud Computing technology for smart city applications, which include latency, traffic congestion, lower throughput, and greater processing time. Some of these drawbacks, however, can be mitigated or even avoided by defining FC in such a way that it moves data processing towards the edge of the network, where data need to be quickly analyzed and decisions made. There is presently little existing work about FC platforms for smart city applications. Nevertheless, we will discuss some related work in the remainder of this section. Very recently,~\cite{9bruneo2016stack4things} designed an OpenStack platform using FC, Stack4Things, to enable smart cities to meet scalability and low latency requirements.
A service-oriented FC architecture~\cite{201dubey2015Fog}, the ``Fog Data,'' aims to reduce Cloud storage and delays in data transmission for telehealth applications by utilizing on-site data processing. To achieve this, the proposed model is designed in three tiers. First, raw data are gathered via wearable sensors and ambient services. Next, a Fog computer is responsible for preliminary data processing and filtering. Finally, a Cloud center conducts a secondary analysis of necessary data. The authors carried out validation of the Fog data for two case studies involving a speech disorder and an ECG.

Summarizing, the previous works reveal how the IoT and FC concepts can help realize the smart city vision and overcome the difficulties associated with remote Cloud data centers. However, none of these works show how we can choose and manage the available and appropriate resources through FNs to further reduce the delays and energy consumption caused by transferring data to far-away FNs. This problem is the main motivation for the multi-tiered framework we propose in section~\ref{sec:4}.
\begin{figure}[!htb]
	\centering
	\includegraphics[width=\textwidth]{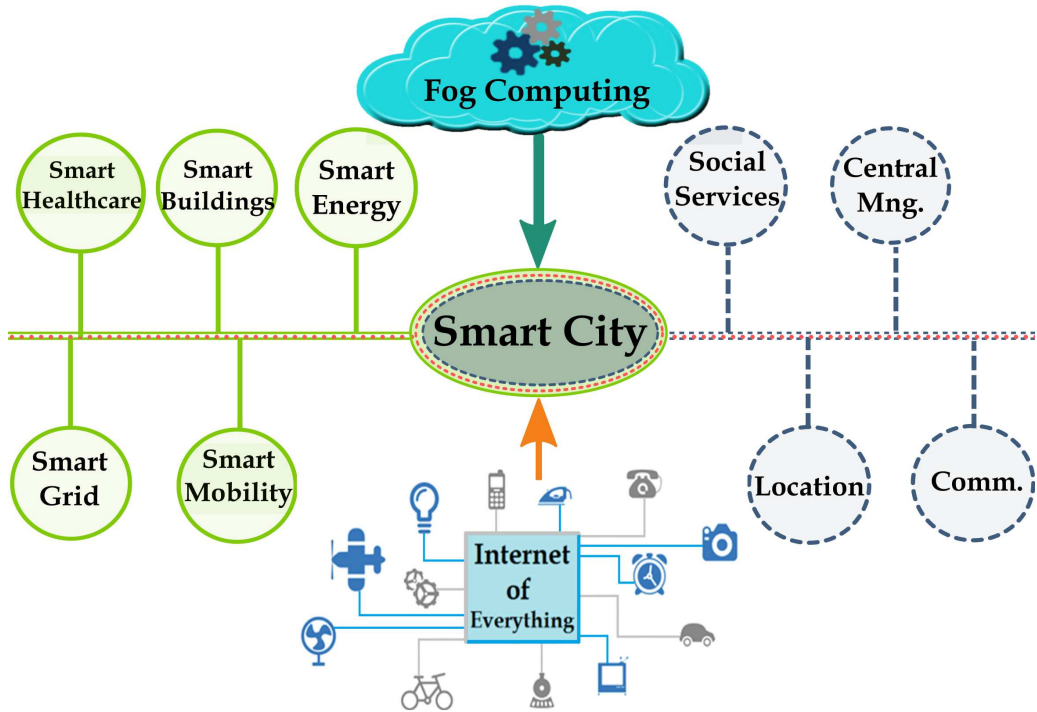}
	\caption{High-level view of the smart city model. Comm:=Communication; Mng:=Management.}
	\label{fig:fig1}
\end{figure}
\section{Model Overview}
\label{sec:3}
This section aims to provide an overview of the proposed model in Fig.~\ref{fig:fig1}. The main objective of the model is to show how smart city components and services can communicate with each other and FC. To this end, the smart city is comprised of several heterogeneous components that serve numerous requests coming from various devices in order to provide these devices with the ability to access different available technologies (e.g., 3G/4G-cellular, WiFi, ZigBee). The devices are connected via the Internet (labeled as the Internet of Everything; see the components at the bottom of Fig.~\ref{fig:fig1}). Without loss of generality, Fig.~\ref{fig:fig1} presents a high-level view of the smart city. The devices in the smart city environment use different services (see the circular components in Fig.~\ref{fig:fig1}), such as smart mobility, a smart grid, smart energy, and so on. These services are used to meet some on-demand requirements that are served from communication devices (micro- and macro-cell objects).

\begin{figure*}[!htb]
	\centering
	\includegraphics[width=\textwidth]{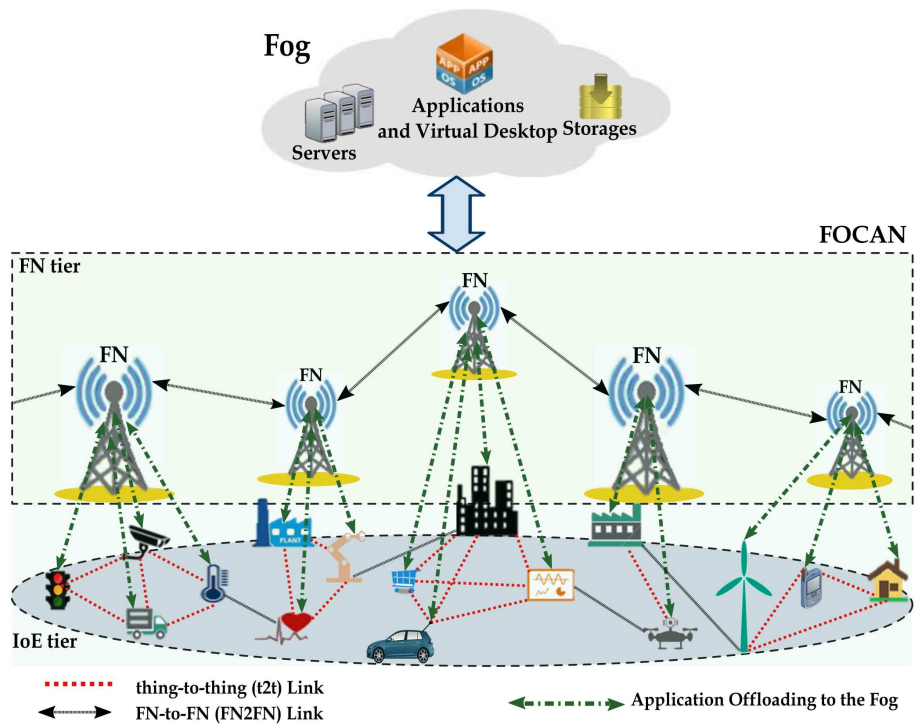}
	\caption{Deployment of Fog-supported smart city architecture. FN:=Fog Node; FOCAN:= Fog Computing Architecture Network.}
	\label{fig:fig2}
\end{figure*}
\section{Fog-supported Smart City Architecture}\label{sec:4}
We propose an architecture for connecting the FNs and the IoE. In future smart cities, technologies will need to be applied in a distributed manner, covering each other in response to users' real-time demands, in order to provide low-latency and high-performance computing for services. These activities will facilitate the residents' quality of life and improve the efficiency of services to meet their needs. Applying FC as a paradigm on top of IoE systems facilitates user services and enables low-latency, high-speed data processing. Motivated by this consideration, we introduce a multi-tiered communication architecture as shown in Fig.~\ref{fig:fig2}, which provides an effective solution for hosting BD applications in the smart cities of the future. Specifically, the proposed architecture, FOCAN (Fog Computing Architecture Network), is comprised of two tiers: (i) the IoE tier, represented with the gray ellipse in the lower part of Fig.~\ref{fig:fig2}, is supported by several heterogeneous devices that are connected to each other and also to the FNs, for communication, and (ii) the FN tier, which covers the incoming traffic from the IoE tier and processes/transfers the data to decrease the latency that is required and thus satisfy the users' service demands.

In the following subsections, we describe the responsibility of each tier and also its relation to the other tier.

\subsection{The IoE Tier}
In this tier, end users can apply any application, anytime, anywhere, and without limitations. In order to preserve the IoE's main function, it is essential to cluster the devices based on their locations. This helps manage the incoming traffic a way that minimizes overhead in terms of time, throughput, and energy consumption. It is possible to serve IoE applications with acceptable latency and throughput when the IoE navigates between heterogeneous hardware and software services if this tier is well integrated and processes IoE data in real time, which increases the proportion of workloads in data centers.

The \textit{things} or \textit{devices} that are the IoE-tier components (see the things inside the IOT tier in Fig.~\ref{fig:fig2}) use TCP/IP-based peer-to-peer (P2P) communication. They can communicate directly with each other via P2P communications when they are near to each other. Otherwise, when things are beyond the range of P2P communications (i.e., beyond the range of Bluetooth, ZigBee, or WiFi communications), they can utilize an FN.

\subsection{The FN Tier}

The Fog-supported smart city provides quality of service (QoS) guarantees to support services. Each FN is composed of a number of physical servers that are interconnected with a wired/wireless access network covering a limited area of radius $R_a(n)$ (i.e., the coverage area for the $n$th FN). The FNs function as small-sized virtualized networked data centers that enable the deployment of Fog services with various types of hardware, such as multicore processors, with fixed hardware resources that can be configured, connected by different network technologies (wired and wireless), and aggregated and abstracted to be viewed as a single logical entity. Each FN serves as a cluster of proximate things that preprocesses and analyzes the real-time data near to the users who are generating the data, facilitating collaboration and proximate social interactions between things (IoT devices), distributed and dynamically.

In addition, the FN provides entry points to a radio access network serving as a wireless communication network technique that is configured to be allotted a single destination (unicast) in the range of the communication. Protocols have been developed to support concurrent data transmissions of the same packet or message to multiple destinations, or data packet broadcasts to all destinations (within a given cell, served by a given service provider, etc.).

From the structural point of view, the FN platform can include a local database that can store applications that are not actively being used in its memory. It can use several retrieval policies to access the data in its buffer in order to decrease the processing time for IoT applications. As part of the Fog-supported smart city architecture, this layer is designed in a way that is capable of implementing social IoT applications (SIoTs) --- that is, FNs can optimize IoE deployment, improving latency, bandwidth, reliability, and security in IoE networks. To this end, FNs can communicate with each other to process data and transfer the required data to the other FNs.

\subsection{Data Communications on FOCAN}
In SIoTs, the data transfer size is rapidly growing. This has led to a decrease in processing speed and the need to retrieve the data from storage and increase the network bandwidth. The data may be obtained from sensor devices, other things (IoT devices), the web, or local storage. The data undergo preprocessing (integrating, filtering, and cleaning) according to the rules that are used for data manipulation.

Instead of sending all data to the Cloud, an FN (an edge device) performs a preliminary analysis and sends an abstract of the metadata to the Cloud. In FOCAN, we consider SIoTs using FCs as an emerging paradigm to allow the splitting of real-time data processing~\cite{9bonomi2012fog} in such a way as to enable the mobility of the end users who are supporting the IoE applications. The FNs in the FN tier provide computing plus networking storage to support the application services for the connected things. Officially, the communications between the components in FOCAN in Fig.~\ref{fig:fig2} are classified as:

\begin{itemize}
	\item \textbf{Primary Communication/Interprimary Communications:} These are local wireless communications in which local devices (things) with processing and sensing capabilities --- such as touch-screen devices, sensors, laptops, and computers --- construct a local P2P inter-thing network to support the wireless communication. For example, WiFi provides primary communication with medium-distance coverage (green dashed lines in the IoE tier in Fig.~\ref{fig:fig2}), and Bluetooth and ZigBee provide interprimary communication, with short-distance coverages between things/devices (red dashed lines in the IoE tier in Fig.~\ref{fig:fig2}) guaranteeing the communication by TCP/IP connections.
    \item \textbf{Secondary Communications:} These are wired/wireless communication between two FNs ($FN2FN$) (see the FN tier in Fig.~\ref{fig:fig2}). It is obvious these should include end-to-end TCP/IP connections such as \textit{IEEE802.11/15} for wireless connections and Cat 5 or 6, or optic fiber, for wired connections.    
\end{itemize}

The primary and interprimary communications are close enough to be supported by a local wireless connection, while the secondary communications are physically or geographically dispersed by more than the range of the local wireless connections. For $t2t$ communications, we have two types of communication: \textit{direct} and \textit{indirect}. Put simply, direct communications are like primary and interprimary communications, whereas indirect communications almost cover the secondary communications. Each FN can communicate with another FN using the direct hopping system. This helps the FOCAN reduce the transfer of data requested by the things between the FNs and thus avoid congestion and jitter and decrease the latency in the network. Furthermore, FNs support secondary types of communication. In a secondary communication, connections are configured to support one or more connections. FNs also guarantee multicasting, which is a transmission of data packets to a given group of destinations that can be performed in a number of ways within wireless communication systems.

\begin{table}[!htpb]
\centering
\caption{FOCAN communication characteristics.}
\label{tab1}
\small
\begin{tabular}{|l|
>{\columncolor[HTML]{F9F9EB}}l |
>{\columncolor[HTML]{F9F9EB}}l |
>{\columncolor[HTML]{F9F9EB}}l |}
\cline{2-4}
\multicolumn{1}{l|}{}                                                                             & \cellcolor[HTML]{57C7E7}\textbf{Primary}                  & \cellcolor[HTML]{57C7E7}\textbf{Interprimary}                   & \cellcolor[HTML]{57C7E7}\textbf{Secondary}                   \\ \hline
\cellcolor[HTML]{E8E8AB}\textbf{Architecture}                                                      & centralized                                               & centralized                                                      & distributed                                                  \\ \hline
\cellcolor[HTML]{E8E8AB}\textbf{QoS}                                                               & high                                                      & high                                                             & very high                                                      \\ \hline
\cellcolor[HTML]{E8E8AB}\textbf{Access medium}                                                     & fixed/wireless                                              & fixed/wireless                                                     & fixed                                                          \\ \hline
\cellcolor[HTML]{E8E8AB}\textbf{Technologies}                                                      & \begin{tabular}[c]{@{}c@{}}WiFi/3G/\\ 4G-LTE\end{tabular} & \begin{tabular}[c]{@{}c@{}}WiFi/\\ Bluetooth/Zigbee\end{tabular} & \begin{tabular}[c]{@{}c@{}}WiFi/3G/\\ 4G-LTE/5G\end{tabular} \\ \hline\hline
\cellcolor[HTML]{E8E8AB}\textbf{Mobility}                                                          & yes                                                       & yes                                                              & no                                                           \\ \hline
\cellcolor[HTML]{E8E8AB}\textbf{Heterogeneity}                                                     & yes                                                       & yes                                                              & yes                                                           \\ \hline
\cellcolor[HTML]{E8E8AB}\textbf{Bandwidth}                                                         & medium                                                    & high                                                             & low                                                          \\ \hline
\cellcolor[HTML]{E8E8AB}\textbf{Latency}                                                           & low                                                       & very low                                                         & low                                                          \\ \hline
\cellcolor[HTML]{E8E8AB}\textbf{Delay Jitter}                                                      & very low                                                  & very low                                                         & low                                                          \\ \hline
\cellcolor[HTML]{E8E8AB}\textbf{\begin{tabular}[l]{@{}l@{}}Stream \\ applications\end{tabular}}    & yes                                                       & yes                                                              & yes                                                          \\ \hline
\cellcolor[HTML]{E8E8AB}\textbf{\begin{tabular}[l]{@{}l@{}}Pervasive \\ applications\end{tabular}} & yes                                                       & yes                                                              & yes                                                           \\ \hline
\cellcolor[HTML]{E8E8AB}\textbf{Storage}                                                           & yes                                                       & no                                                               & yes                                                          \\ \hline
\cellcolor[HTML]{E8E8AB}\textbf{Protocols}                                                         & \multicolumn{3}{c|}{\cellcolor[HTML]{F9F9EB}CDMA/TDMA/FDMA/OFDM/GSM}                                                                                                                        \\ \hline
\end{tabular}
\end{table}

Table~\ref{tab1} presents the FOCAN communication characteristics. It is important to emphasize that the secondary communications are distributed all over the FOCAN, while the primary and interprimary communications are confined to the range of the FNs. The primary communications increase the QoS much better than the two other types of communications, due to the use of high-speed communications and low latency for the service processing. The mobility of things has an important influence on their communications, which can be handled with primary and interprimary communications. Moreover, FOCAN supports device and FN heterogeneity, and stream-type/pervasive applications and secondary communications, due to the use of FNs in these communications, have much less jitter compared to the other communication types.       
\begin{figure}[!htb]
	\centering
	\includegraphics[width=0.8\textwidth]{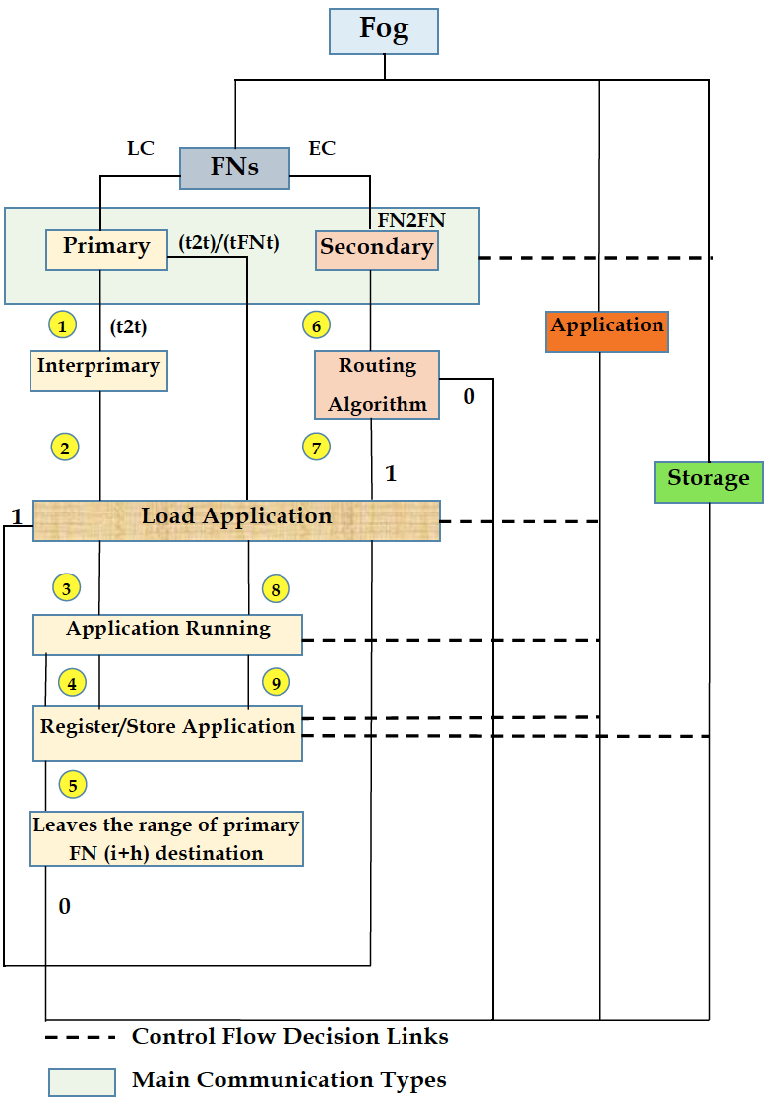}
	\caption{FOCAN model diagram.}
	\label{fig:fig3}
\end{figure}

Figure~\ref{fig:fig3} shows the FOCAN flow diagram. The Fog component in Fig.~\ref{fig:fig2} comprises several servers, applications, and storage devices (as shown in the uppermost Cloud-shaped component named ``Fog'' in Fig.~\ref{fig:fig2}). Therefore, it can be divided into \textit{three} groups of blocks (see the dark blue, red, and green box components in Fig.~\ref{fig:fig3}). In the FOCAN architecture, each FN device covers a server in the Fog component.  The FN block in Fig.~\ref{fig:fig2} can be subdivided into primary and secondary communication types, as shown in the light green box under the FN main block. The primary and secondary communication blocks use wired/wireless communications for transferring data between the first and second tiers of FOCAN. In this framework, the servers (FNs) are allocated among several applications that can be dynamically shrunk or expanded based on the service's requirements, real-time demands, and the available resources in the FN tier, which can be handled in a distributed way by the FNs. The FN communications can be classified into Local Communications (LCs) and External Communications (ECs). An LC is interpreted as a primary communication and an EC is interpreted as a secondary communication. Primary communications can be subtyped as the interprimary box, which expresses the $t2t$ relationships. To follow the flow of this type of communication, we begin by authenticating and authorizing the applications that each thing needs to utilize (see the dashed-line control-flow decision between the Application box and the Load Application box in Fig. \ref{fig:fig3}), over the congestion-aware safe communication link in the FOCAN IoE tier. Then we run the applications and save the processed applications in the FN's storage using the dashed-line control link to the storage component in Fig.~\ref{fig:fig3}. This same process is conducted for the $tFNt$ communications. Furthermore, for the EC communication type linked to the secondary communication subblock, $FN2FN$, we need to find the proper destination FN from a source FN when one of the things covered by the source FN wants to communicate with a thing in the destination FN. Thus, we need to find the routing path (see the routing algorithm block in Fig.~\ref{fig:fig3}) by identifying the hops for transferring the required information between these two things. The first step here is to find a traversing path to identify the hops. We assume that the path from the source FN to the destination FN is labeled ``1'' and the return path is labeled ``0''. After finding the routing path and the direction, if it is labeled ``1'', the request is sent from the source node $FN(i)$ to the destination node. We do steps 1 and 2 and load the application into the FN buffer, traverse $h$ hops to reach the destination $FN(i+h)$, and then execute steps 3 and 4, saving/registering the application in the storage of the destination FN. After that, step 5 occurs, and we leave $FN(i+h)$ and return to the source node $FN(i)$ (following label ``0'' in Fig.~\ref{fig:fig3}, which is connected to the ``Routing Algorithm'' box), and then we repeat steps 6 to 9 from the ``Routing Algorithm'' box for the applications. Note that this will happen if an FN is communicating with another FN ($FN2FN$), so it is essential to use routing algorithms for the connection.   
\begin{figure}[!htb]
	\centering
	\includegraphics[width=\textwidth]{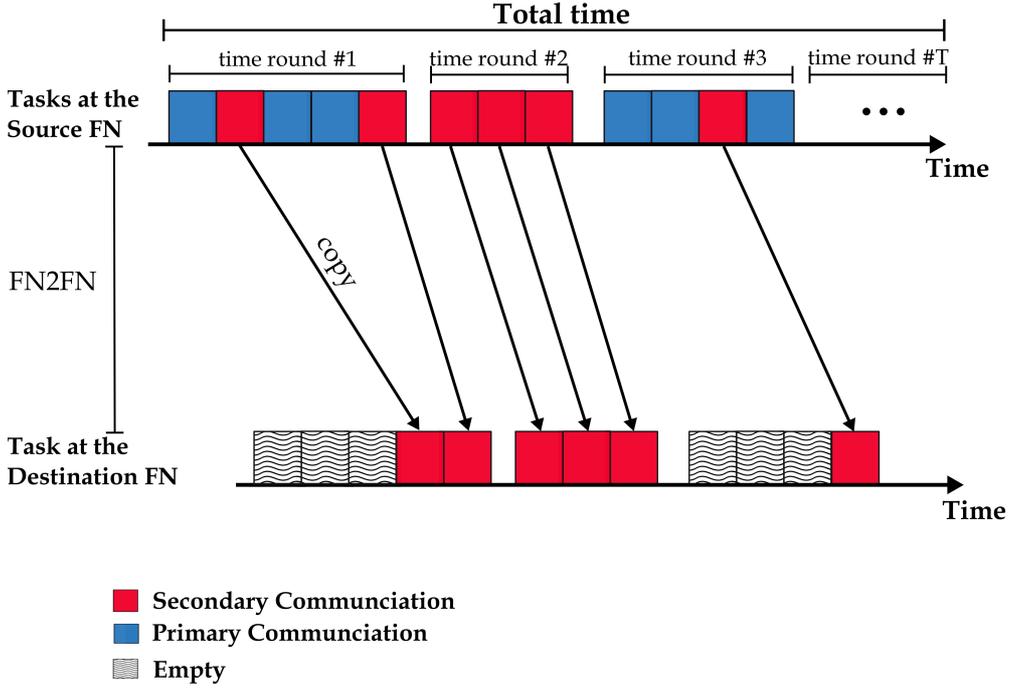}
	\caption{Time chart for the secondary communications between FNs.}
	\label{fig:fig4}
\end{figure}  

To understand how the FN buffer is processed, we present the time chart for the data packet scheduling technique between the source and destination FNs in Fig.~\ref{fig:fig4}, following~\cite{clark2005live}. We modify~\cite{clark2005live} in such a way that the application's packets are transferred from the source FN to the destination FN. These packets include secondary communication tasks and primary communication tasks, distinguished in the figure by color. This time chart also adopts the Time-Division Multiple Access (TDMA) routing algorithm recently presented in \cite{naranjo2016p} as a viable solution for $FN2FN$ communications. In accordance with the TDMA technique presented in \cite{naranjo2016p}, we use the time rounds of the packets and the overall TTL, $T$, for transferring a set of packets to the identified destination FN, $FN(i+h)$. In the next section, we present a holistic case study and evaluate the FOCAN with several cases. 
  
\section{Performance Evaluation and Validation} \label{sec:5}
In this section, we present real-time-based scenarios, which are heightened in smart cities.

\subsection {Simulation Setup}
To evaluate and compare the performances of the FOCAN platform, we conducted some numerical simulations. The experiments for our proposed platform were implemented on an iFogSim simulator, which provides real-time scenarios for Fog-based networks in a smart city~\cite{ifogsim}. In addition, we use the same network setting as~\cite{naranjo2016p}. Here, the incoming traffic, called tasks, refers to the web-based application demands of two types of resources (CPU and intra-/inter-network) that are needed for executing the I/O traffic on the things in the smart city. 

The static FNs are deployed in the city based on physical proximity (e.g., 10 meters, 30 meters, 3 feet)~\cite{ifogsim}; each FN serves a spatial cluster of the corresponding radius value, acting as a service point for the things currently in the served cluster. The FN comprises a heterogeneous multicore server that can simultaneously run multiple instructions time on AMD, a Phenom II X6 1090T BE 6-core x86 architecture processor, equipped with 3.2 GHz and 6 GB of RAM for each core. These settings have great gains in processor performance by virtue of increasing the operating frequency, which allows higher performance at lower energy. A wired Giga Ethernet switch connects the FNs, and each FN has primary and secondary communications. Each FN consumes electric power to process the incoming traffic. 
The tasks are performed on the FM cores at  the processing frequency 10 Mb/s 
with a maximum rate of 2.5 Gb/s. 
We assume that the tasks are uniformly allocated to each core. In order to evaluate the CPU and network power cost for each FN, we use the CPU and network power formulas of~\cite{AMD} and set the maximum and idle CPU power at 195 and 105 watts, respectively. 
Also, we fixed the delays for $t2t$, $tFNt$, and $FN2FN$ communications at 2, 4, and 6 ms, respectively~\cite{ifogsim}. 
In addition, the configuration's average round trip time for the wireless and wired communication were set at 0.5 ms and 10 ms, 
respectively~\cite{AMD,naranjo2016p}. The simulations were carried out for a period of 1000 s.

\subsection {Simulation Results}
\begin{figure}[!htb]
	\centering
	\subfloat[Normalized sampled trace of an I/O traffic flow from an enterprise data center, using Messenger.]{\includegraphics[width=0.5\textwidth]{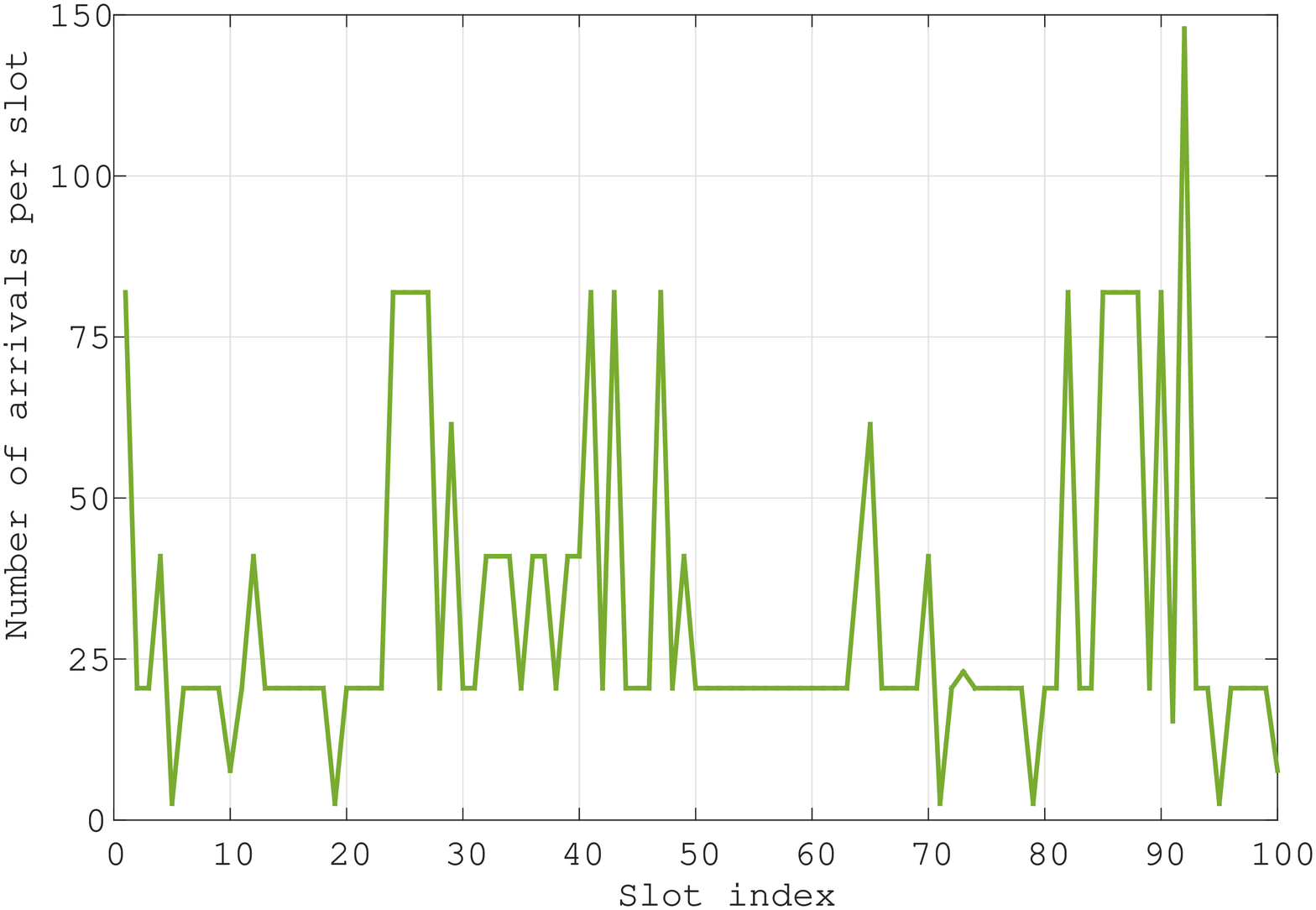}\label{fig:w}}\hfill
	\subfloat[Per-communication and per-round average power consumption.]{\includegraphics[width=0.5\textwidth]{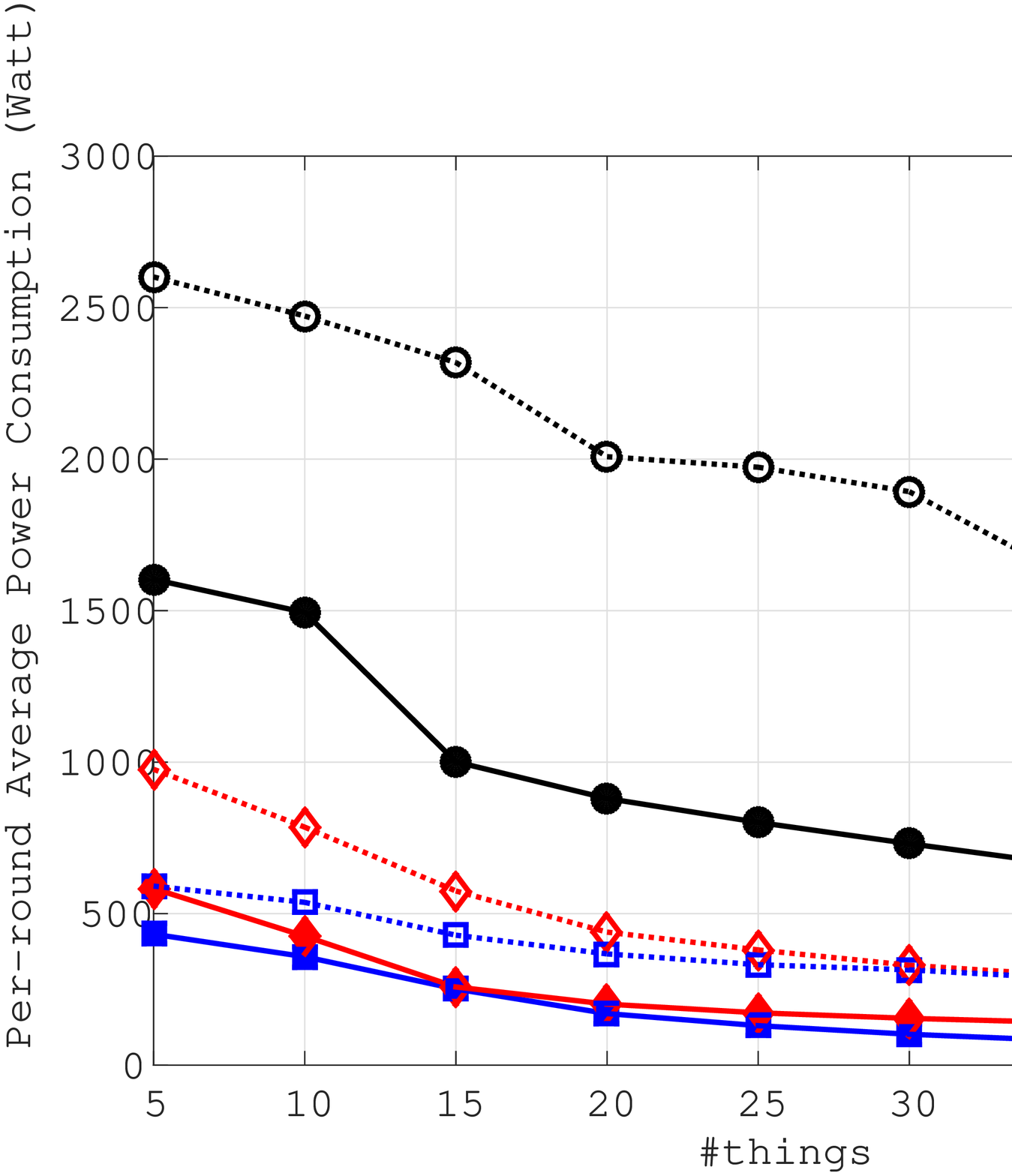}\label{fig:f5a}}
	\caption{The simulation results for FOCAN for various communication rates.}
	\label{fig:f5}
\end{figure}

We validated the FOCAN in terms of the average overall energy consumption (average of the CPU and network costs) per processing time by comparing the FOCAN and the D2D in~\cite{li2014exploring} in terms of various communication costs (see Fig.~\ref{fig:f5}). We ran our proposed solution and evaluated the resulting average total consumed power for each FN under a web-based application (i.e., MSN Messenger input arrival) and results are shown in Fig.~\ref{fig:w}; this is a normalized traffic trace that reports the I/O real-workload traffic flow to the different types of communications. Note that the FNs can be integrated or distributed all over the model presented in Fig.~\ref{fig:fig2}.
\begin{figure}[!htb]
	\centering
	\subfloat[Per-communication and per-round average power consumption for D2D under the settings in \cite{li2014exploring}.]{\includegraphics[width=0.5\textwidth]{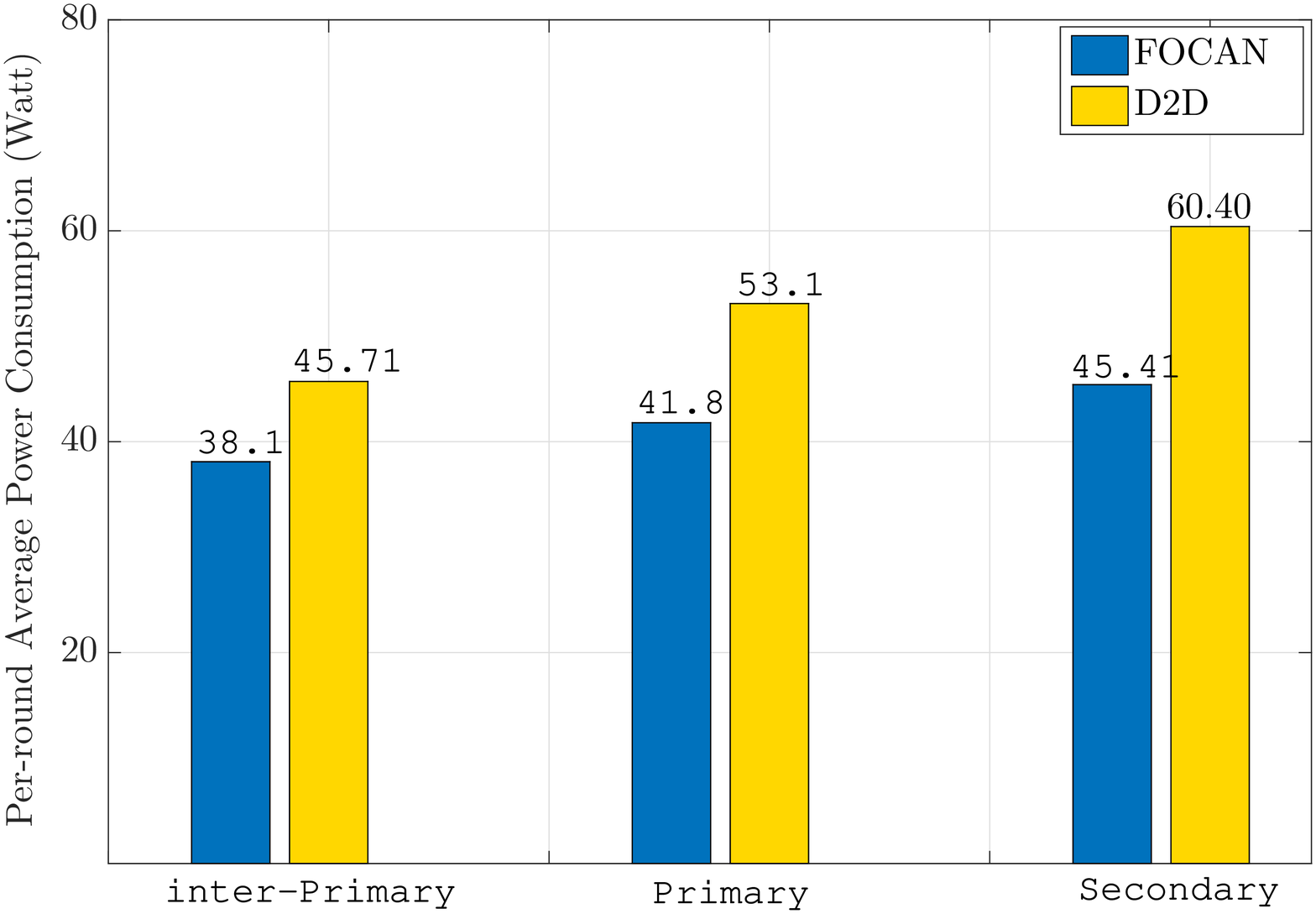}\label{fig:f5b}}
	\subfloat[FNs' average total power consumption for the Messenger data.]{\includegraphics[width=0.5\textwidth]{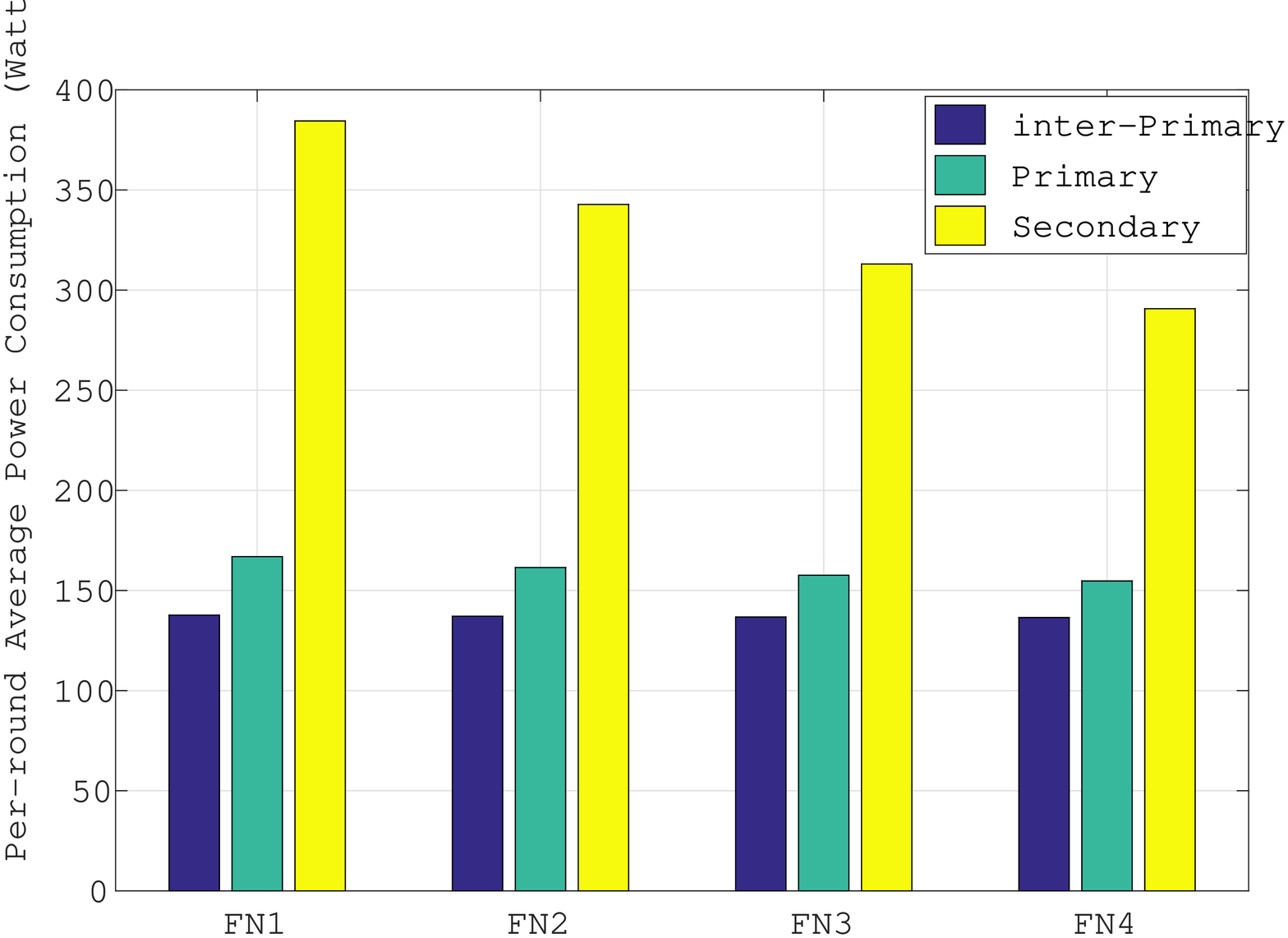}\label{fig:f5c}}
	\caption{The simulation results for FOCAN compared with the D2D approach in~\cite{li2014exploring}.}
	\label{fig:f55}
\end{figure}
The numerical results in Fig.~\ref{fig:f5b} for the FOCAN framework report the average energy and time per round; we simulated the underlying $t2t$, $t2FN$, and $FN2FN$ wired/wireless links that carried out the types of communication activities. For comparison purposes, we simulated a $D2D$ platform that works, according to~\cite{li2014exploring}, uses \textit{IEEE802.11b} with single-hop D2D links FOR $t2t$ connections, with the traffic flows transported by all simulated TCP/IP connections. Fig.~\ref{fig:f5b} shows that the FOCAN platform is more power efficient than the D2D platform; with the corresponding average per-connection power, the D2D connections increase the power consumption because it is affected by the fading and path loss and the average number of TCP time-out events and packet retransmissions, which is confirmed in Fig.~\ref{fig:f5b}. Finally, Fig.~\ref{fig:f5c} shows the average power consumption for an evaluation of the interprimary, primary, and secondary communications. 

\section {Open Issues and Challenges}\label{sec:6}
Academic research has developed a wide variety of techniques and technologies to capture, curate, analyze, and visualize BD. An integrated system includes network infrastructure services, education information services, and learning services. The leading benefit is that new knowledge is obtained and higher-order thinking skills are facilitated when each student generates a unique data track where data can be inserted, processed, and analyzed~\cite{selinger2013education}. Today, more and more classrooms are becoming ``open'' through voice-, video-, and text-based collaboration, and teachers have a wide range of multimodal resources to enhance their teaching. In the field of computer science, the challenge is to develop different forms of innovative education that can be adjusted for large numbers of students around the world, engage students with different interests, and that can carry out a new curriculum that shows the fundamental changes in computing technology~\cite{selinger2013education}.
We also need to support the integration and collaboration of different governments to improve business decisions through BD analytics. To achieve this, governments need to publish new policies for handling data that specify rowners' and producers' restrictions. This legislation will be useful for managing the quality of the data and processing the real-time analytics of huge streaming data volumes. Also, the new legislation can help in developing a massively scalable scheme for enabling the visualization of information from thousands of real-time sources, encompassing application development built on Hoop, stream computing, and data warehousing.
\section{Conclusion and Future Directions}\label{sec:7}

Considering the thousands of smart city applications that are running on numerous things, as well as the emergence of FC to cover such applications by running at
the edge of the Internet to meet the requirements of scalability, energy awareness, and low latency, we have designed a framework called \textit{FOCAN} for managing things' applications. FOCAN can be  classified as a computation- and communication-efficient structure and scalable routing algorithm that minimizes the average power consumption of FNs. Noteworthy features of the developed FOCAN include: (i) it minimizes the energy consumed by the overall FOCAN platform for computing, intra-Fog communication, and wired/wireless transmission over thing-aware TCP/IP connections; (ii) it subsumes the IoE device communications over FNs under three categories: interprimary, primary, and secondary, to arrange traffic and manage tasks across the FNs, and (iii) the flow diagram for the systems' incoming tasks and suggested routing algorithms shows how the data can be transferred to the corresponding things to guarantee the Fog applications. A quantitative analysis demonstrates that FOCAN allows effective management of small areas within an urban region; hence, it provides scalable energy-aware Fog-supported application management. 

This work can be extended in several directions of potential interest. For example, it can be extended to cover 5G management in light of the huge number of things that use stream applications, e.g., online video chatting, or sets of things that are playing online games with other sets in other regions of the city. To do this, we need to add real-time data processing solutions together with using Mobile Edge Computing techniques to make robust frameworks.

 
\section*{Acknowledgments} \label{sec:9}
Mauro Conti is supported by a Marie Curie Fellowship funded by the European Commission (agreement PCIG11-GA-2012-321980). This work is also partially supported by the EU TagItSmart! Project (agreement H2020-ICT30-2015-688061), the EU-India REACH Project (agreement ICI+/2014/342-896), and by the projects ``Physical-Layer Security for Wireless Communication", and ``Content Centric Networking: Security and Privacy Issues" funded by the University of Padua. This work is partially supported by the grant n. 2017-166478 (3696) from Cisco University Research Program Fund and Silicon Valley Community Foundation and Melbourne-Chindia Cloud Computing (MC$^3$) Research Network.

\bibliographystyle{elsarticle-num}
\bibliography{bibliography}
\end{document}